\documentclass{nature-arxiv}

\usepackage{acronym}
\usepackage{amsmath}
\usepackage{amssymb}
\usepackage{color}
\usepackage{hyperref}
\usepackage{graphicx}

\newcommand{\chieff}{\chi_\mathrm{eff}}
\newcommand{\dd}{\mathrm{d}}

\newcommand{\OOneSigmaIsoAligned}{2.4}
\newcommand{\OOneOddsIsoAligned}{0.015}
\newcommand{\OTwoSigmaIsoAlignedMin}{2.4}
\newcommand{\OTwoOddsIsoAlignedMin}{0.016}

\newcommand{\plotone}[1]{\includegraphics[width=\columnwidth]{#1}}

\begin{document}

\acrodef{BBH}{binary black hole}
\acrodef{BH}{black hole}
\acrodef{EM}{electromagnetic}
\acrodef{GW}{gravitational wave}
\acrodef{O1}{first observing}
\acrodef{PE}{parameter estimation}

\title{Distinguishing Spin-Aligned and Isotropic Black Hole
  Populations With Gravitational Waves}

\author{Will M. Farr$^1$, Simon Stevenson$^{1,3}$, M. Coleman Miller$^2$,
  Ilya Mandel$^{1,3}$, Ben Farr$^4$ \& Alberto Vecchio$^1$}

\maketitle

\begin{affiliations}
\item Birmingham Institute for Gravitational Wave Astronomy and
  School of Physics and Astronomy, University of Birmingham,
  Birmingham, B15 2TT, United Kingdom
\item Department of Astronomy and Joint Space-Science
  Institute, University of Maryland, College Park, MD 20742--2421,
  United States
\item Kavli Institute for Theoretical Physics, Santa Barbara, CA 93106
\item Enrico Fermi Institute and Kavli Institute for Cosmological
  Physics, University of Chicago, Chicago, IL 60637, United States
\end{affiliations}

\begin{abstract}
  The first direct detections of gravitational
  waves\cite{2016PhRvL.116f1102A,2016PhRvL.116x1103A,O1-BBH,PhysRevLett.118.221101}
  from merging binary black holes open a unique window into the binary
  black hole formation environment.  One promising environmental
  signature is the angular distribution of the black hole spins;
  systems formed through dynamical interactions among already-compact
  objects are expected to have isotropic spin
  orientations\cite{SigurdssonHernquist:1993,PZMcMillan:2000,Rodriguez:2015,Stone:2016,2016ApJ...832L...2R}
  whereas binaries formed from pairs of stars born together are more
  likely to have spins preferentially aligned with the binary orbital
  angular
  momentum\cite{TutukovYungelson:1993,2016Natur.534..512B,Stevenson:2017,MandeldeMink:2016,Marchant:2016}.
  We consider existing gravitational wave measurements of the binary
  effective spin, the best-measured combination of spin
  parameters\cite{O1-BBH,PhysRevLett.118.221101}, in the four likely
  binary black hole detections GW150914, LVT151012, GW151226, and
  GW170104.  If binary black hole spin magnitudes extend to high
  values we show that the data exhibit a $\OOneSigmaIsoAligned\sigma$
  ($\OOneOddsIsoAligned$ odds ratio\footnote{An odds ratio of $r$ with
    $r \ll 1$ is equivalent to $x \sigma$ with
    $x = \Phi^{-1}\left( 1 - r/2 \right)$, where $\Phi$ is the unit
    normal CDF.})  preference for an isotropic angular distribution
  over an aligned one.  By considering the effect of 10 additional
  detections\cite{2016ApJ...833L...1A}, we show that such an augmented
  data set would enable in most cases a preference stronger than
  $5\sigma$ ($2.9 \times 10^{-7}$ odds ratio).  The existing
  preference for either an isotropic spin distribution or low spin
  magnitudes for the observed systems will be confirmed (or
  overturned) confidently in the near future.
\end{abstract}

\acresetall{}

Following the detection of a merging binary black hole system,
parameter estimation tools compare model gravitational waveforms
against the observed data to obtain a posterior distribution on the
parameters that describe the compact binary source.  The spin
parameter with the largest effect on waveforms, and a correspondingly
tight constraint from the data\cite{O1-BBH}, is a mass-weighted
combination of the components of the dimensionless spin vectors of the
two black holes that are aligned with the orbital axis, the
``effective spin,'' $-1 < \chieff < 1$ (see Methods Section
\ref{methsec:chieff-spin-magnitude}).

\begin{figure}
  \plotone{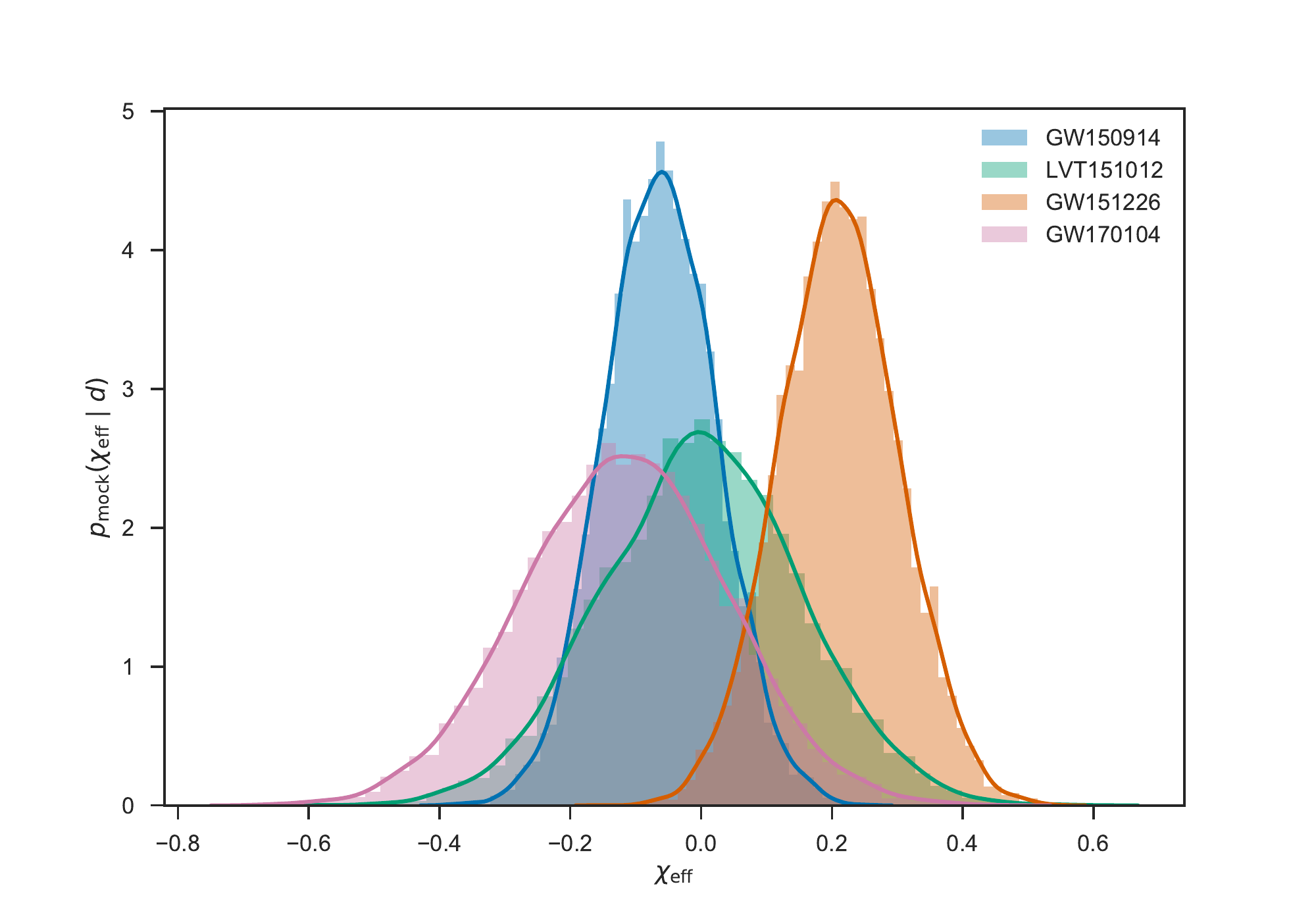}
  \caption{\label{fig:O1-posteriors} \textbf{Approximate posteriors on
      $\chieff$ from the Advanced LIGO O1 and GW170104
      observations\cite{O1-BBH,PhysRevLett.118.221101}.}  We
    approximate the posteriors reported using Gaussians with the same
    median and 90\% credible interval.  It is notable that none of the
    $\chieff$ posteriors support high \ac{BH} spin magnitudes with
    aligned spins, suggested by observations of stellar-mass black
    holes in X-ray binaries\cite{2015PhR...548....1M}.}
\end{figure}

Figure \ref{fig:O1-posteriors} shows an approximation to the posterior
inferred on $\chieff$ for the four likely \ac{GW} detections GW150914,
GW151226, GW170104, and LVT151012 from Advanced LIGO's first and
second observing runs (O1 and O2)\cite{O1-BBH,PhysRevLett.118.221101}.
Because samples drawn from the posterior on $\chieff$ are not publicly
released at this time, we have approximated the posterior as a
Gaussian distribution with the same mean and 90\% credible interval,
truncated to $-1 < \chieff < 1$.  None of the $\chieff$ posteriors are
consistent with two black holes with large aligned spins,
$\chi_{1,2} \gtrsim 0.5$; this contrasts with the large spins inferred
for the majority of black holes in X-ray binaries with claimed spin
measurements\cite{2015PhR...548....1M} (see below).  The analysis here
is relatively insensitive to the precise details of the posterior
distributions; other conclusions are more sensitive.  In particular,
our Gaussian approximation does permit $\chieff = 0$ for GW151226
while the true posterior rules this out at high
confidence\cite{2016PhRvL.116x1103A,O1-BBH}.

Small values of $\chieff$ as exhibited in these systems can result
from either intrinsically small spins or larger spins whose direction
is mis-aligned with the orbital angular momentum of the binary (i.e.\
spin vectors with small $z$-components).  Mis-alignment is capable of
producing \emph{negative} values of $\chieff$, however, whereas
aligned spins will always have $\chieff \geq 0$.  This difference
provides strong discriminating power between the two angular
distributions, even without good information about the magnitude
distribution; to the extent that data favour negative $\chieff$ they
weigh heavily against aligned models.  To quantify the degree of
support for these two alternate explanations of small $\chieff$ values
in the merging binary black hole population, we compared the Bayesian
evidence for various simple models of the spin population using the
\ac{GW} data set.

\begin{figure}
  \plotone{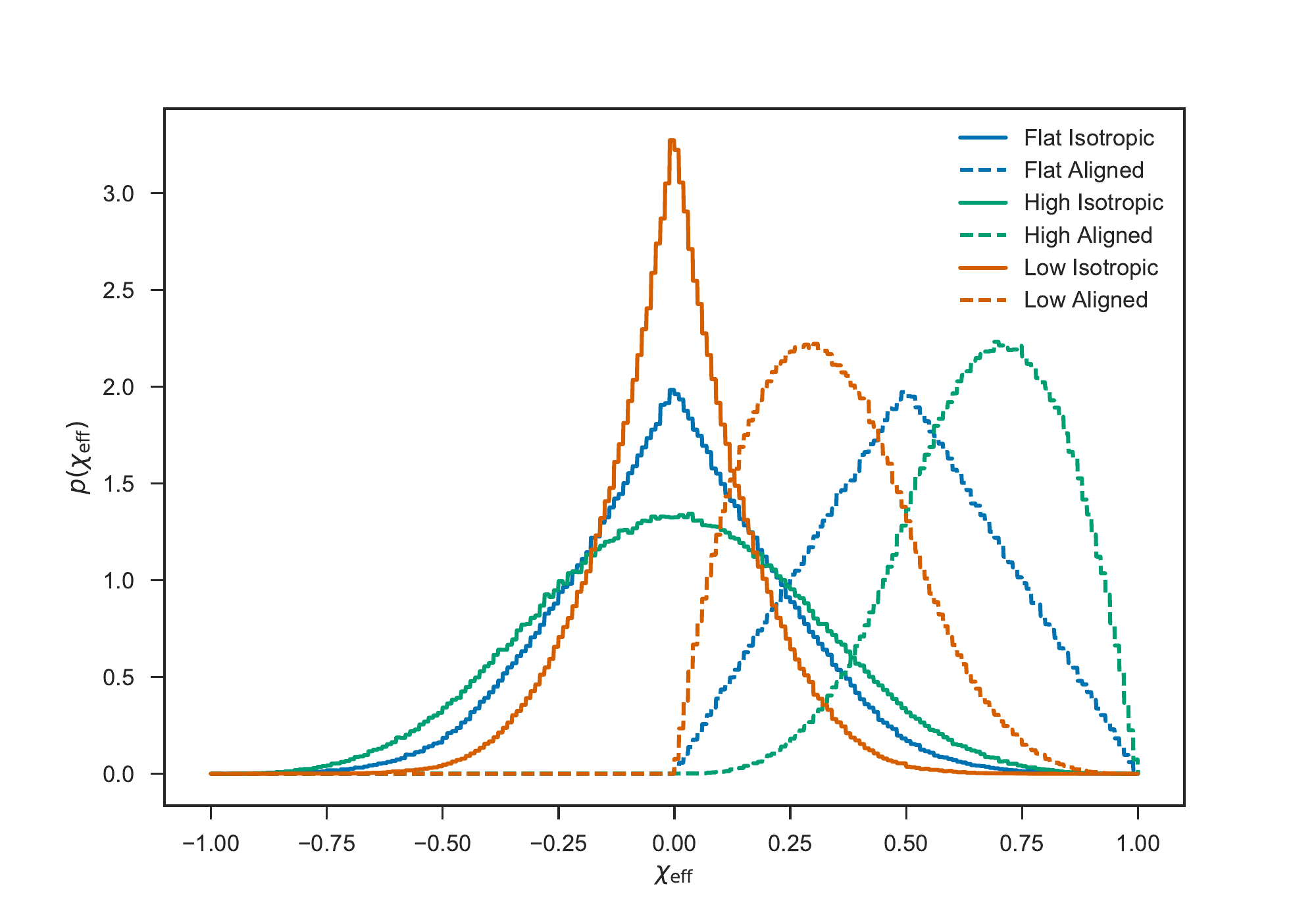}
  \caption{\label{fig:chieff-distribution-models} \textbf{The models for the
    population distribution of $\chieff$ considered in this paper.}  In
    all models we assume that the binary mass ratio
    $q \equiv m_1/m_2 = 1$ and that the distribution of spin vectors
    is the same for each component.  The ``flat'' (blue lines),
    ``high,'' (green lines), and ``low'' (red lines) magnitude
    distributions are defined in Eq.\ \eqref{eq:magnitude-dists}.
    Solid lines give the $\chieff$ distribution under the assumption
    that the orientations of the spins are isotropic; dashed lines
    give the distribution under the assumption that both objects'
    spins are aligned with the orbital angular momentum.  The
    isotropic distributions are readily distinguished from the aligned
    distributions by the production of negative $\chieff$ values,
    while the distinction between the three models for the spin
    magnitude distribution is less sharp.}
\end{figure}

Each of our models for the merging binary black hole spin population
assumes that the merging black holes are of equal mass (this is
marginally consistent with the
observations\cite{O1-BBH,PhysRevLett.118.221101}, and the $\chieff$
distribution is not sensitive to the mass ratio---see Methods Section
\ref{sec:mass-ratio}).  We assume that the population spin
distribution factorises into a distribution for the spin magnitude $a$
and a distribution for the spin angles.  Finally, we assume that the
distribution of spins is common to each component in a merging binary
(the distributions of spin for each component in the binary could
differ systematically due to different formation histories).  Choosing
one of three magnitude distributions (see Methods Section
\ref{methsec:chieff-spin-magnitude}), ``low'' (mean $a = 0.33$,
standard deviation $0.24$), ``flat'' (mean $a = 0.5$, standard
deviation $0.29$), ``high'' (mean $a = 0.67$, standard deviation
$0.24$) and pairing with an isotropic angular distribution or a
distribution that generates perfect alignment yields six different
models for the $\chieff$ distribution.  These models are shown in
Figure \ref{fig:chieff-distribution-models}.

These magnitude distributions are not meant to represent any
particular physical model, but rather to capture our uncertainty about
the spin magnitude distribution; neither observations nor population
synthesis codes can at this point authoritatively suggest \emph{any}
particular spin distribution\cite{2015PhR...548....1M}.  Our models,
however, allow us to see how sensitive the $\chieff$ distribution is
to spin alignment given uncertainties about the spin magnitudes.

\begin{figure}
  \plotone{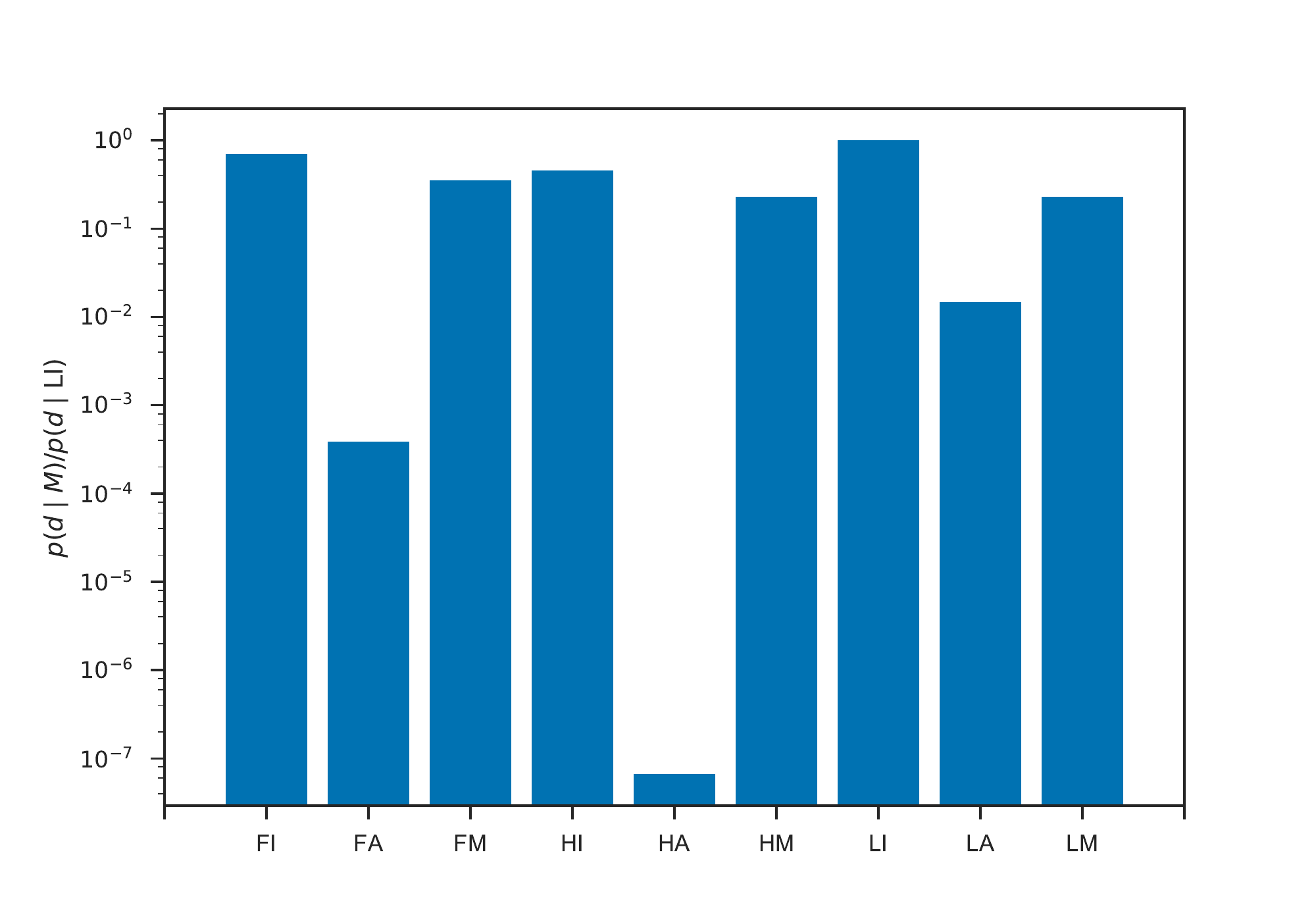}
  \caption{\textbf{Odds ratios among our models using the
      approximations to the posteriors on $\chieff$ from the O1 and GW170104
      observations shown in Figure \ref{fig:O1-posteriors}.}  The flat
    (``F''), high (``H''), and low (``L'') spin magnitude
    distributions (see Eq.\ \eqref{eq:magnitude-dists}) are paired
    with isotropic (``I'') and aligned (``A'') angular distributions,
    as well as a mixture model of the two (``$\mathrm{M}$'').  The
    most-favoured models have the ``low'' distribution of spin
    magnitudes.  The odds ratio between the best aligned and best
    isotropic models is $\OOneOddsIsoAligned$, or
    $\OOneSigmaIsoAligned\sigma$.  For all magnitude distributions the
    pure-isotropic models are preferred over the mixture models;
    correspondingly, the posterior on the mixture fraction peaks at
    100\% isotropic.}
  \label{fig:O1-odds}
\end{figure}

We fit hierarchical models of the three existing LIGO O1 and GW170104
observations using these six different, zero-parameter population
distributions (see Methods Section \ref{sec:hierarchical}).  We also
fit three mixture models for the population, where the angular
distribution is a weighted sum of the isotropic and aligned
distributions.  The evidence, or marginal likelihood, for each of the
models is shown in Figure \ref{fig:O1-odds}.  For all three magnitude
distributions, the mixture models' posterior on the mixing fraction
peaks at 100\% isotropic.  Not surprisingly, given the small $\chieff$
values in the three detected systems, the most-favoured model among
those with an isotropic angular distribution has the ``low'' magnitude
distribution; the most favoured model among those with an aligned
distribution also has the ``low'' magnitude distribution.  The odds
ratio between the ``low'' aligned and ``low'' isotropic models is
$\OOneOddsIsoAligned$, or $\OOneSigmaIsoAligned\sigma$; thus the data
favour isotropic spins among our suite of models.  While the data
favour spin amplitude distributions with small spin magnitudes, note
that a model with all binary black hole systems having zero spin is
ruled out by the GW151226 measurements, which bound at least one black
hole to have spin magnitude $\geq 0.2$ at 99\%
credibility\cite{2016PhRvL.116x1103A}.

\begin{figure}
  \plotone{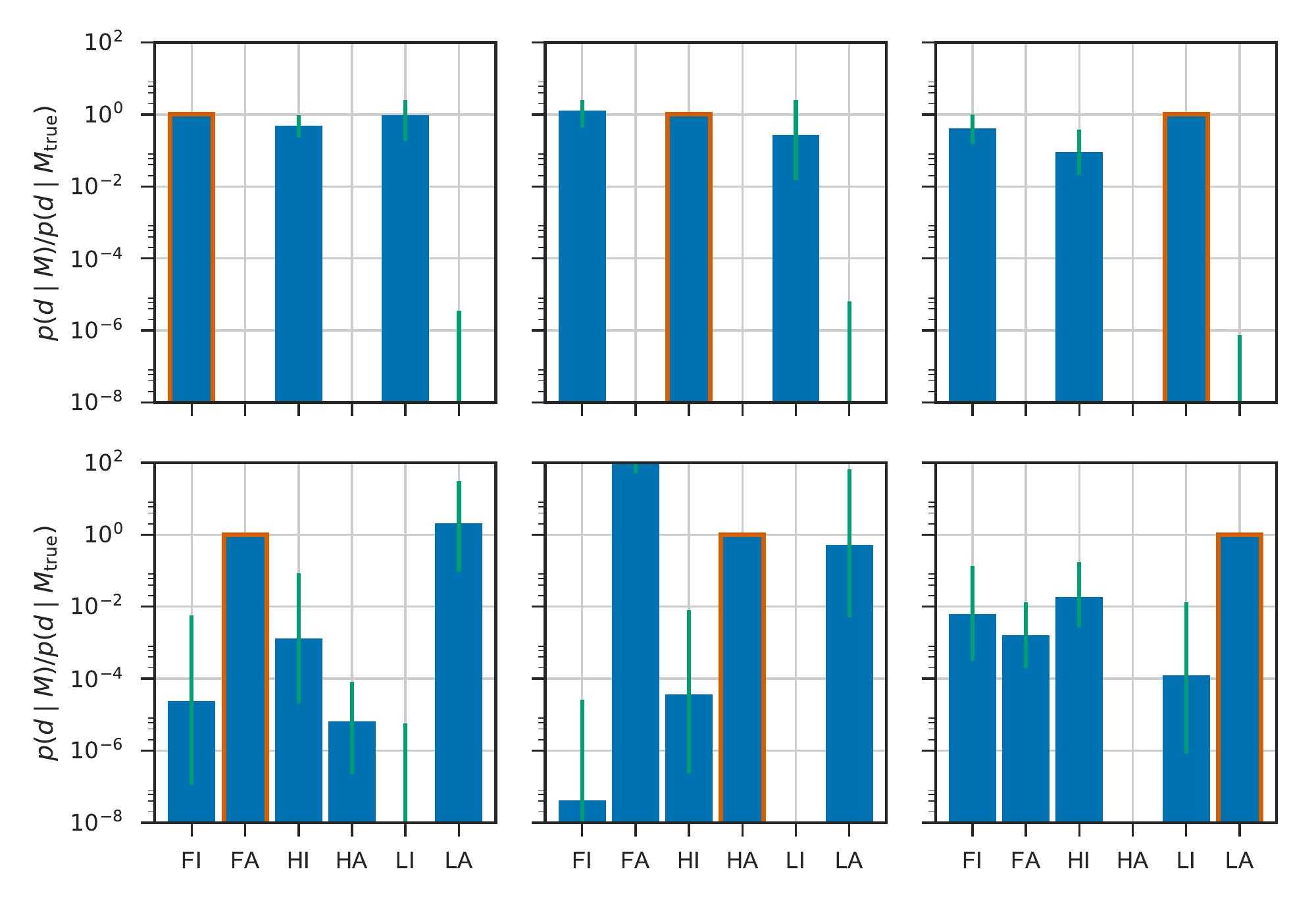}
  \caption{\label{fig:O2-predictions} \textbf{Distribution of odds
      ratios predicted with 10 additional observations above the four
      discussed above.}  Each panel corresponds to additional
    observations drawn from one of the $\chieff$ distribution models.
    The model from which the additional observations are drawn is
    outlined in red.  The height of the blue bar gives the median odds
    ratio relative to the model from which the additional observations
    are drawn; the green line gives the 68\% ($1 \sigma$) symmetric
    interval of odds ratios over 1000 separate draws from the model
    distribution.  The closest median ratio between the most-favoured
    isotropic model and the most-favoured aligned model is
    $\OTwoOddsIsoAlignedMin$, corresponding to
    $\OTwoSigmaIsoAlignedMin\sigma$ preference for the correct angular
    distribution; most models result in more than $5\sigma$ preference
    for the correct angular distribution.  Because the four existing
    observations are included in each data set the ``correct'' model
    is not necessarily preferred over the others, particularly when
    that model uses the ``high'' magnitude distribution, which is
    strongly dis-favoured from the O1 and GW170104 observations
    alone.}
\end{figure}

Estimates of the rate of binary black hole coalescences give a
reasonable chance of 10 additional binary black hole detections in the
next three years\cite{O1-BBH,2016ApJ...833L...1A}.  Assuming 10
additional detections drawn from each of our six zero-parameter models
in addition to the four existing detections from O1 and GW170104, with
observational uncertainties drawn randomly from the three Gaussian
widths used to approximate the $\chieff$ posteriors in Figure
\ref{fig:O1-posteriors}\footnote{The measurement uncertainty in
  $\chieff$ depends on the other parameters of the merging binary
  black hole system, particularly on the mass ratio.  Our assumption
  about future observational uncertainties is appropriate if the
  parameters of the three detected events are representative of the
  parameters of future detections. See Methods Section
  \ref{sec:chi-eff-precision} for further discussion.}, we find the
odds ratios shown in Figure \ref{fig:O2-predictions}.  We find that
most scenarios with an additional 10 detections allow the simulated
angular distribution to be inferred with greater than $5\sigma$
($2.9 \times 10^{-7}$ odds) credibility. In the most pessimistic case
the distinction is typically $\OTwoSigmaIsoAlignedMin\sigma$
($\OTwoOddsIsoAlignedMin$ odds ratio).  While such future detections
should permit a confident distinction between \emph{angular}
distributions, we would remain much less certain about the
\emph{magnitude} distribution among the three options considered here
until we have a larger number of observations.  

Most of our resolving power for the spin angular distribution is a
result of the fact that our ``aligned'' models cannot produce
$\chieff < 0$ (see Figure \ref{fig:chieff-distribution-models}).  If
spins are intrinsically very small, with $a \lesssim 0.2$, then it is
no longer possible to resolve the negative effective spin with a small
number of observations (see Methods Section \ref{sec:smallspins}).  As
noted below, however, spins observed in X-ray binaries are typically
large.  Additionally, models which do not permit \emph{some} spins
with $\chieff \gtrsim 0.1$ are ruled out by the GW151226
observations\cite{2016PhRvL.116x1103A}.  An ``aligned'' model with
spin magnitudes from our ``flat'' distribution but permitting spin
vectors oriented anti-parallel to the orbital angular momentum
(leading to the possibility of positive \emph{or} negative $\chieff$)
can only be distinguished from an isotropic true population at
$\sim 3 \sigma$ after 10--20 observations\cite{2017CQGra..34cLT01V};
our flat aligned model can be distinguished from such a population at
better than $5\sigma$ (odds $< 10^{-8}$) after 10 observations,
emphasizing the information content of the bound $\chieff > 0$ for our
aligned models.

Observational data on spin magnitudes in black hole systems is
sparse\cite{2015PhR...548....1M}.  Most of the systems studied are
low-mass X-ray binaries rather than the high-mass X-ray binaries that
are likely to be the progenitors of double black hole binaries.  In
addition, there are substantial systematic errors that can complicate
these analyses\cite{2015PhR...548....1M} and selection effects could
yield a biased distribution. Nonetheless, if we take the reported spin
magnitudes as representative then we find that there is a preference
for high spins; for example, 14 of the 19 systems with reported spins
have dimensionless spin parameters in excess of 0.5.  It is usually
argued that the masses and spin parameters of stellar-mass black holes
are unlikely to be altered significantly by
accretion\cite{1999MNRAS.305..654K}, but this may not be true for all
systems\cite{2015ApJ...800...17F}.  Thus the current spin parameters
are probably close to their values upon core collapse, at least in
high-mass X-ray binaries.  However, the specific processes involved in
the production of black hole binaries from isolated binaries could
alter the spin magnitude distribution of those holes relative to the
X-ray binary systems; for example, close tidal interactions could spin
up the core, or stripping of the envelope could reduce the available
angular
momentum\cite{2016MNRAS.462..844K,2017arXiv170200885Z,2017arXiv170203952H}.

The spin directions in isolated binary black
holes\cite{TutukovYungelson:1993,2016Natur.534..512B,Stevenson:2017,MandeldeMink:2016,Marchant:2016}
are usually expected to be preferentially aligned.  Despite observed
spin-orbit misalignments in massive stellar
binaries\cite{Albrecht:2009}, mass transfer and tidal interactions
will tend to realign the binary.  On the other hand, there is some
evidence of spin-orbit misalignment in black hole X-ray
binaries\cite{Martin:2008b,MorningstarMiller:2014}.  This is
consistent with the expectation that a supernova natal kick (if any)
can change the orbital plane and misalign the
binary\cite{2000ApJ...541..319K}; the supernova can also tilt the spin
angle\cite{2011ApJ...742...81F}.  Evolutionary processes, such as
wind-driven mass loss and post-collapse fallback, can couple the spin
magnitude and direction distributions, contrary to our simplified
assumptions.  A small misalignment at wide separation can also evolve
to a more significant misalignment in component spins as the binary
spirals in through \ac{GW} emission\cite{2015PhRvD..92f4016G}, but
$\chieff$ is approximately conserved through this evolution.

The spin directions of binary black holes formed dynamically through
interactions in dense stellar
environments\cite{SigurdssonHernquist:1993,PZMcMillan:2000,Rodriguez:2015,Stone:2016}
are expected to be isotropic given the absence of a preferred
direction\cite{2016ApJ...832L...2R} and the persistence of an
isotropic distribution through post-Newtonian
evolution\cite{2004PhRvD..70l4020S,2007ApJ...661L.147B}.

\begin{addendum}
\item We thank Richard O'Shaughnessy, Christopher Berry, Davide
  Gerosa, and Salvatore Vitale for discussions and comments on this
  work.  WF, SS, IM and AV were supported in part by the STFC.  MCM
  acknowledges support of the University of Birmingham Institute for
  Advanced Study Distinguished Visiting Fellows program.  SS and IM
  acknowledge support from the National Science Foundation under Grant
  No. NSF PHY11-25915.
\item[Supplementary Information] Supplementary Information is linked
  to the online version of the paper at
  \url{http://www.nature.com/nature}.
\item[Author Contributions] All authors contributed to the work
  presented in this paper.
\item[Author Information] Reprints and permissions information is
  available at \url{http://www.nature.com/reprints}.  The authors
  declare no competing financial interests.  Correspondence and
  requests for materials should be addressed to
  \href{mailto:w.farr@bham.ac.uk}{w.farr@bham.ac.uk}.
\end{addendum}

\begin{methods}

% Reset the Figure Counter
\setcounter{figure}{0}
\renewcommand{\figurename}{Extended Data Figure}

\section{Code Availability}

This analysis used the Julia language\cite{Julia}, Python libraries
NumPy and SciPy\cite{NumPy,SciPy}, the plotting library
Matplotlib\cite{Matplotlib}, and performed computations in IPython
notebooks\cite{IPython}.  A repository containing the code and
notebooks used for this analysis, together with the \LaTeX{} source
for this document, can be found under an open-source ``MIT'' license
at \url{https://github.com/farr/AlignedVersusIsoSpin}.

\section{Effective Spin and Spin Magnitude Distributions}
\label{methsec:chieff-spin-magnitude}

The effective spin is defined by\cite{2016PhRvL.116x1102A} 
\begin{equation}
  \chieff = \frac{c}{GM} \left( \frac{\vec{S}_1}{m_1} + \frac{\vec{S}_2}{m_2}
  \right) \cdot \frac{\vec{L}}{\left| \vec{L} \right|} \equiv \frac{1}{M} \left( m_1 \chi_1 + m_2 \chi_2 \right),
\end{equation}
where $m_{1,2}$ are the gravitational masses of the more-massive (1)
and less-massive (2) components, $M = m_1 + m_2$ is the total mass,
$\vec{S}_{1,2}$ are the spin angular momentum vectors of the black
holes in the binary, $\vec{L}$ is the orbital angular momentum vector,
assumed to point in the $\hat{z}$ direction, and $\chi_{1,2}$ are the
corresponding dimensionless projections of the individual \ac{BH}
spins.  Because the dimensionless spin parameter,
\begin{equation}
  \label{eq:a-def}
  a_{1,2} = \frac{c}{G m_{1,2}^2} \left|\vec{S}_{1,2} \right|,
\end{equation}
of each black hole is bounded by $0 \leq a_{1,2} < 1$, the projections
along the orbital axis are bounded by $-1 < \chi_{1,2} < 1$, and
$-1 < \chieff < 1$.

We form the population distributions of $\chieff$ shown in Figure
\ref{fig:chieff-distribution-models} by assuming that each black hole
in a binary has a dimensionless spin magnitude drawn from one of three
distributions,
\begin{equation}
  \label{eq:magnitude-dists}
  p(a) = \begin{cases}
    2\left(1-a \right) & \textnormal{``low''} \\
    1 & \textnormal{``flat''} \\
    2 a & \textnormal{``high''}
  \end{cases},
\end{equation}
referred to as ``low,'' ``flat,'' and ``high'' in the text above.
These distributions are shown in Extended Data Figure \ref{fig:pa}.

\begin{figure}
  \includegraphics[width=\columnwidth]{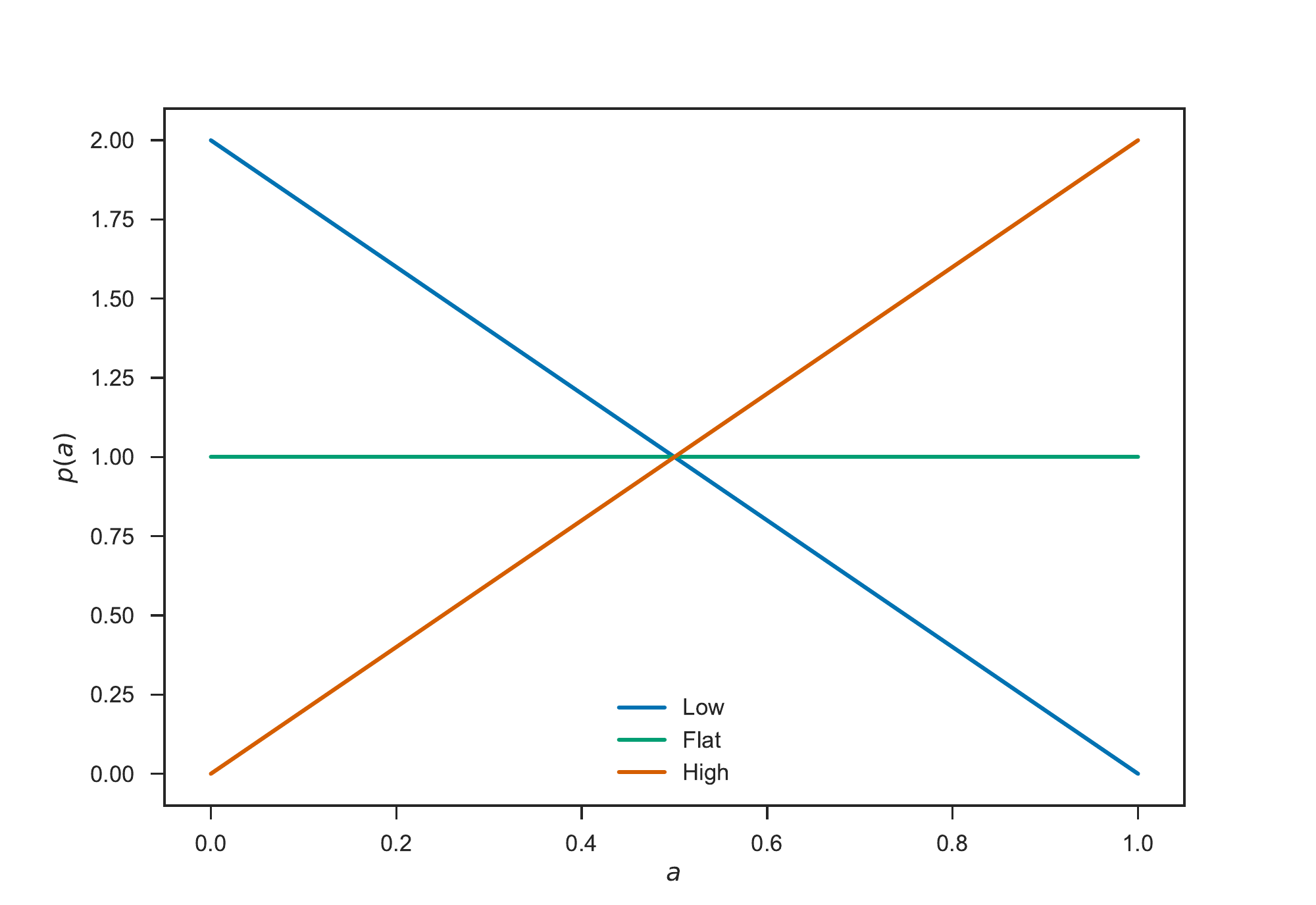}
  \caption{\textbf{Distributions of spin magnitudes.}  See Eq.\
    \eqref{eq:magnitude-dists} for the definition of the low (blue
    line), flat (green line), and high (red line) magnitude
    distributions used here.  The distributions have mean spin $0.33$,
    $0.5$, and $0.67$ and standard deviations $0.24$, $0.29$, and
    $0.24$.}
  \label{fig:pa}
\end{figure}

\section{Mixture model}

While we carried out Bayesian comparisons between isotropic and aligned spin distributions under various assumptions, a preference for one of the considered models over the others does not necessarily indicate that it is the correct model.  All of the considered models could be inaccurate for the actual distribution, especially since all of the considered models are based on a number of additional assumptions, such as decoupled spin magnitude and spin misalignment angle distributions and identical distributions for primary and secondary spins.

We now partly relax the simplified assumptions made earlier by considering the possibility that the true distribution of BBH spin-orbit misalignments observed by LIGO is a mixture of binaries with aligned spins and binaries with isotropic spins.

\begin{figure}
\centering
\includegraphics[width=\textwidth]{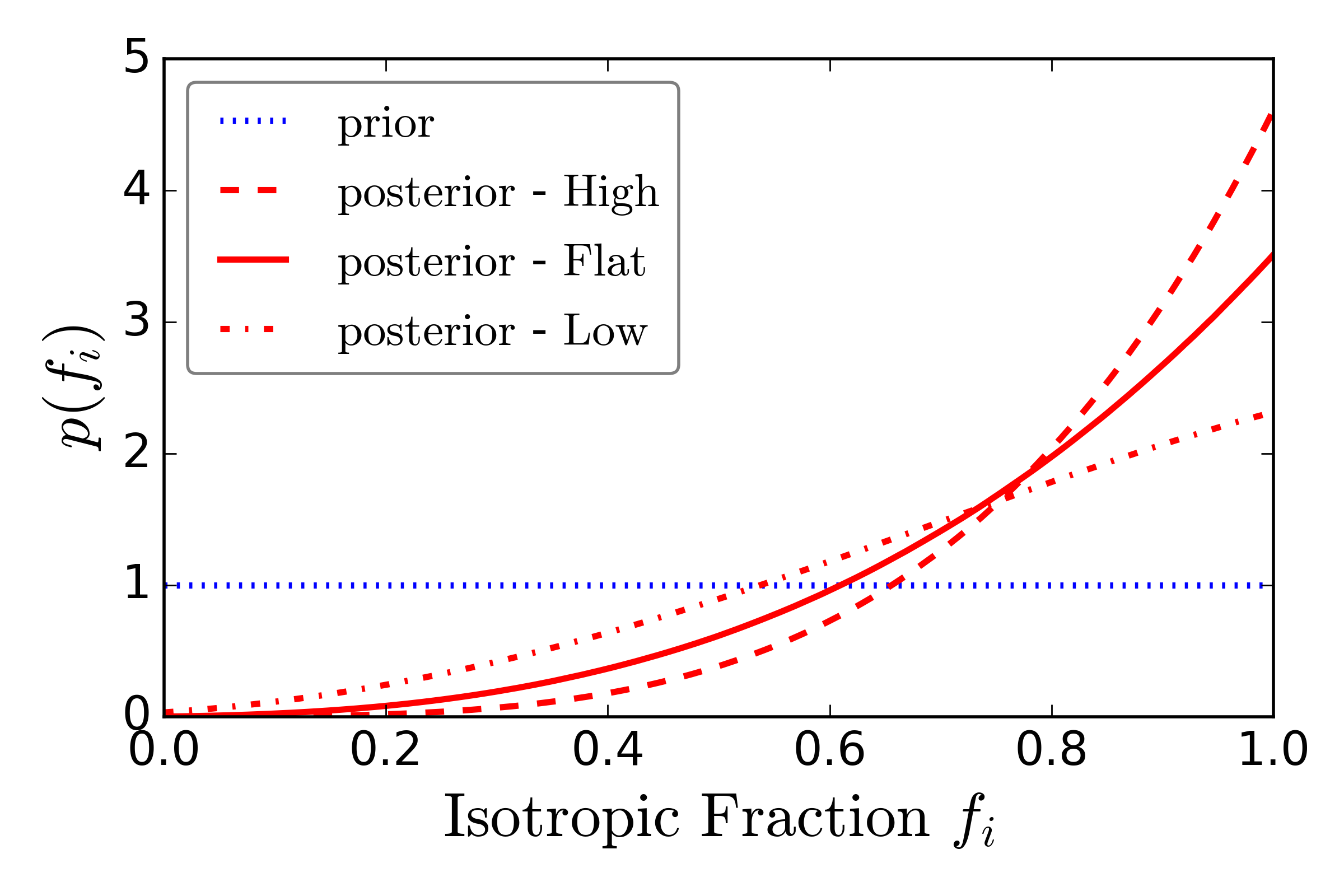}
\caption{\textbf{Fraction of the BBH population coming from an
    isotropic distribution under a mixture model.} The dotted line
  shows the flat prior on the fraction of BBHs coming from an
  isotropic distribution, $f_i$, under the mixture model. The 3 red
  lines show the posterior on $f_i$ after O1 and GW170104 with our
  various assumptions regarding BH spin magnitudes.  The solid line
  shows the posterior assuming that all BHs have their spin magnitude
  drawn from the ``flat'' distribution. The dashed line assumes the
  ``high'' BH spin magnitude distribution $p(a) = 2a$. The dot-dash
  line assumes the ``low'' distribution $p(a) = 2(1-a)$.  We see that
  for a wide range of assumptions regarding BH spin magnitudes, the
  fraction coming from an isotropic distribution $f_i$ peaks at 1.}
\label{fig:mixture_fraction_posterior}
\end{figure}
%

%-- could also use the cumulative posterior
%%
%\begin{figure}
%\centering
%\includegraphics[width=0.45\textwidth]{plots/isotropic_fraction_cumulative_posterior.png}
%\caption{\textbf{Fraction of the BBH population coming from an isotropic distribution} The dotted (blue) line shows the cumulative distribution for the flat prior on the fraction of BBHs coming from an isotropic distribution $f_i$. The solid (red) line shows the cumulative distribution for the posterior after O1, assuming that all BHs have their spin magnitude drawn from a uniform distribution. The dashed line assumes a `increasing' distribution $p(a) = 2a$ for BH spin magnitudes, whilst the dot-dash line assumes a `decreasing' distribution $p(a) = 2(1-a)$. We see that regardless of our assumption regarding BH spin magnitudes, there is a preference for a large fraction coming from an isotropic distribution.}
%\label{fig:mixture_fraction_posterior}
%\end{figure}
%

We fit a mixture model\cite{Stevenson:2017spin} (labelled model 'M' in
Figure~\ref{fig:O1-odds}) where a fraction $f_i$ of BBHs have spins
drawn from an isotropic distribution, whilst a fraction $1 - f_i$ have
their spins aligned with the orbital angular momentum. We assume a
flat prior on the fraction $f_i$. To test the robustness of our
result, we vary the distribution we assume for BH spin magnitude
distributions as with the aligned and isotropic models. We use the
``flat'', ``high'' and ``low'' distributions
(Equation~\ref{eq:magnitude-dists}), assuming all BHs have their spin
magnitude drawn from the same distribution for both the aligned and
isotropic populations. We calculate and plot the posterior on $f_i$
given by Equation~\ref{eq:hierarchical-posterior} ($f_i=\lambda$ in
the derivation) in Extended Data
Figure~\ref{fig:mixture_fraction_posterior}. We find the mean fraction
of BBHs coming from an isotropic distribution is 0.70, 0.77 and 0.81
assuming the ``low'', ``flat'' and ``high'' distributions for spin
magnitudes respectively, compared to the prior mean of 0.5. The lower
90\% limits are 0.38, 0.51 and 0.60 respectively, compared to the
prior of 0.1. In all cases, the posterior peaks at $f_i = 1$. Thus,
for these spin magnitude distributions we find that the current O1 and
GW170104 LIGO observations constrain the majority of BBHs to have
their spins drawn from an isotropic distribution. The evidence ratios
of these mixture models to the isotropic distribution with ``low''
spin magnitudes are 0.43, 0.20 and 0.10 for the ``low'', ``flat'' and
``high'' spin magnitude models. Thus we cannot rule out a mixture with
the current data.  If several different components contribute
significantly to the true spin distribution it may take tens to
hundreds of detections to accurately determined the mixing fraction,
depending on the distribution of spin
magnitudes\cite{2017CQGra..34cLT01V,Stevenson:2017spin}.

\section{Hierarchical Modelling} 
\label{sec:hierarchical}

LIGO measures $\chieff$ better than any other spin parameter, but
still with significant uncertainty\cite{2015PhRvD..91d2003V}, so we
need to properly incorporate measurement uncertainty in our analysis;
thus our analysis must be
\emph{hierarchical}\cite{2010ApJ...725.2166H,2010PhRvD..81h4029M}.  In
a hierarchical analysis, we assume that each event has a true, but
unknown, value of the effective spin, drawn from the population
distribution, which may have some parameters $\lambda$; then the
system is observed, represented by the likelihood function, which
results in a distribution for the true effective spin (and all other
parameters describing the system) consistent with the data.
Combining, the joint posterior on each system's $\chieff^i$ parameters
and the population parameters $\lambda$ implied by a set of
observations each with data $d^i$, is
\begin{equation}
  p\left( \left\{ \chieff^i \right\}, \lambda \mid \left\{ d^i \right\} \right) \propto \left[ \prod_{i=1}^{N_\mathrm{obs}} p\left(d^i \mid \chieff^i \right) p\left( \chieff^i \mid \lambda \right) \right] p\left(\lambda\right).
\end{equation}

The components of this formula are
\begin{itemize}
\item The GW (marginal) likelihood, $p\left(d \mid \chieff\right)$.
  Here we use ``marginal'' because we are (implicitly) integrating
  over all parameters of the signal but $\chieff$.  Note that it is
  the likelihood rather than the posterior that matters for the
  hierarchical analysis; if we are given posterior distributions or
  posterior samples, we need to re-weight to ``remove'' the prior and
  obtain the likelihood.
\item The population distribution for $\chieff$,
  $p\left( \chieff \mid \lambda \right)$.  This function can be
  parameterised by population-level parameters, $\lambda$.  (In the
  cases discussed above, there are no parameters for the population.)
\item The prior on the population-level parameters, $p(\lambda)$.
\end{itemize}
If we do not care about the individual event $\chieff$ parameters, we
can integrate them out, obtaining
\begin{equation}
  p\left( \lambda \mid \left\{ d^i \right\} \right) \propto \left[ \prod_{i=1}^{N_\mathrm{obs}} \int \dd \chieff^i \, p\left(d^i \mid \chieff^i \right) p\left( \chieff^i \mid \lambda \right) \right] p\left(\lambda\right).
\end{equation}
If we are given posterior samples of $\chieff^{ij}$ ($i$ labels the
event, $j$ labels the particular posterior sample) drawn from an
analysis using a prior $p\left( \chieff \right)$, then we can
approximate the integral by a re-weighted average of the population
distribution over the samples (here $p\left( \chieff^{ij} \right)$ is
the prior used to produce the posterior samples):
\begin{equation}
  p\left( \lambda \mid \left\{ d^i \right\} \right) \propto \left[ \prod_{i=1}^{N_\mathrm{obs}} \frac{1}{N_i} \sum_{j=1}^{N_i} \frac{p\left( \chieff^{ij} \mid \lambda \right)}{p\left( \chieff^{ij} \right)} \right] p\left(\lambda\right).
  \label{eq:hierarchical-posterior}
\end{equation}

\subsection{Order of Magnitude Calculation}
\label{sec:om-odds-ratio}

It is possible to estimate at an order-of-magnitude level the rate at
which evidence accumulates in favour of or against the isotropic
models as more systems are detected.  Based on Figure
\ref{fig:chieff-distribution-models}, approximate the isotropic
population $\chieff$ distribution as uniform on
$\chieff \in \left[ -0.25, 0.25 \right]$ and the aligned population
$\chieff$ distribution as uniform on
$\chieff \in \left[0, 0.5\right]$.  Then the odds ratio between the
isotropic and aligned models for each event is approximately
\begin{equation}
  \label{eq:approx-odds}
  \frac{p\left( d \mid I \right)}{p\left( d \mid A \right)} \simeq
  \frac{P\left( -0.25 \leq \chieff \leq 0.25 \right)}{P\left( 0 \leq \chieff \leq 0.5 \right) },
\end{equation}
where $P\left( A \leq \chieff \leq B \right)$ is the posterior
probability (here used to approximate the likelihood) that $\chieff$
is between $A$ and $B$.  Using our approximations to the $\chieff$
posteriors described above, this gives an odds ratio of $5$ in favour
of the isotropic models, which is about a factor of two smaller than
the ratio in the more careful calculation described above.  This is a
satisfactory answer at an order-of-magnitude level.

If the true distribution is isotropic and follows this simple model,
and our measurement uncertainties on $\chieff$ are $\simeq 0.1$, then
the geometric mean of each subsequent measurement's contribution to
the overall odds is $\sim 3$.  After ten additional events, then, the
odds ratio becomes $5 \times 3^{10} \simeq 3 \times 10^{5}$, or
$4.6 \sigma$, consistent with the results of the more detailed
calculation described above.  If the true distribution of spins
becomes half as wide ($\chieff \in [-0.125, 0.125]$ for isotropic and
$\chieff \in [0, 0.25]$ for aligned spins), with the same
uncertainties, then the existing odds ratio becomes $1.08$, and each
subsequent event drawn from the isotropic distribution contributes on
average a factor of $1.6$.  In this case, after 10 additional events,
the odds ratio becomes $150$, or $2.7\sigma$.  With small spin
magnitudes, our angular resolving power vanishes, as discussed in more
detail in Methods Section \ref{sec:smallspins}.

\subsection{Accumulation of evidence}

In Table~\ref{tab:accumulation} we show how the evidence for an isotropic distribution increases when including: only the 2 confirmed events---GW150914 and GW151226---from O1; all O1 events (including LVT151012); and all 4 likely binary black hole mergers, including GW170104. 

\begin{table}
\begin{centering}
\begin{tabular}{ c |  c  |  c  | c  }
  \hline 
  \hline
Events & $\sigma_{I/A}$ &  $\sigma_{I/A}$ & $\sigma_{I/A}$ \\ 
 & ``Low'' & ``Flat'' & ``High'' \\ \hline
GW150914 and GW151226 &  1.3 & 2.2 & 3.7 \\ %\hline
All O1 events & 1.7 & 2.7 & 4.4 \\ %\hline
All O1 events and GW170104 & \textbf{2.4} & 3.6 & 5.4 \\ \hline
\end{tabular}
\caption[]{Significance $\sigma_{I/A}$ of the odds ratio between the isotropic and aligned models using just GW150914 and GW151226, all 3 O1 events, and all 4 currently observed events (including GW170104). The number in bold is the result we quote in the main text.}
\label{tab:accumulation}
\end{centering}
\end{table}

\section{Effect of small spin magnitudes}
\label{sec:smallspins}

In the main text we considered three models for BH spin magnitudes:
``low'', ``flat'' and ``high''. These were intended to capture some of
the uncertainty regarding the BH spin magnitude distribution.  We may
remain observationally uncertain about the spin magnitude distribution
until we have $\mathcal{O}(100)$
observations\cite{2017arXiv170306869F,2017arXiv170306223G}.

Here we extend the ``low'' model as:
\begin{equation}
p(a) \propto (1 - a)^{\alpha}
\label{eq:lowspinalpha}
\end{equation}

When $\alpha = 0$, this recovers the ``flat'' distribution, whilst
$\alpha = 1$ recovers the ``low'' distribution. For higher values of
$\alpha$, this distribution becomes more peaked towards
$a = 0$.

\begin{figure}
\centering
\includegraphics[width=\textwidth]{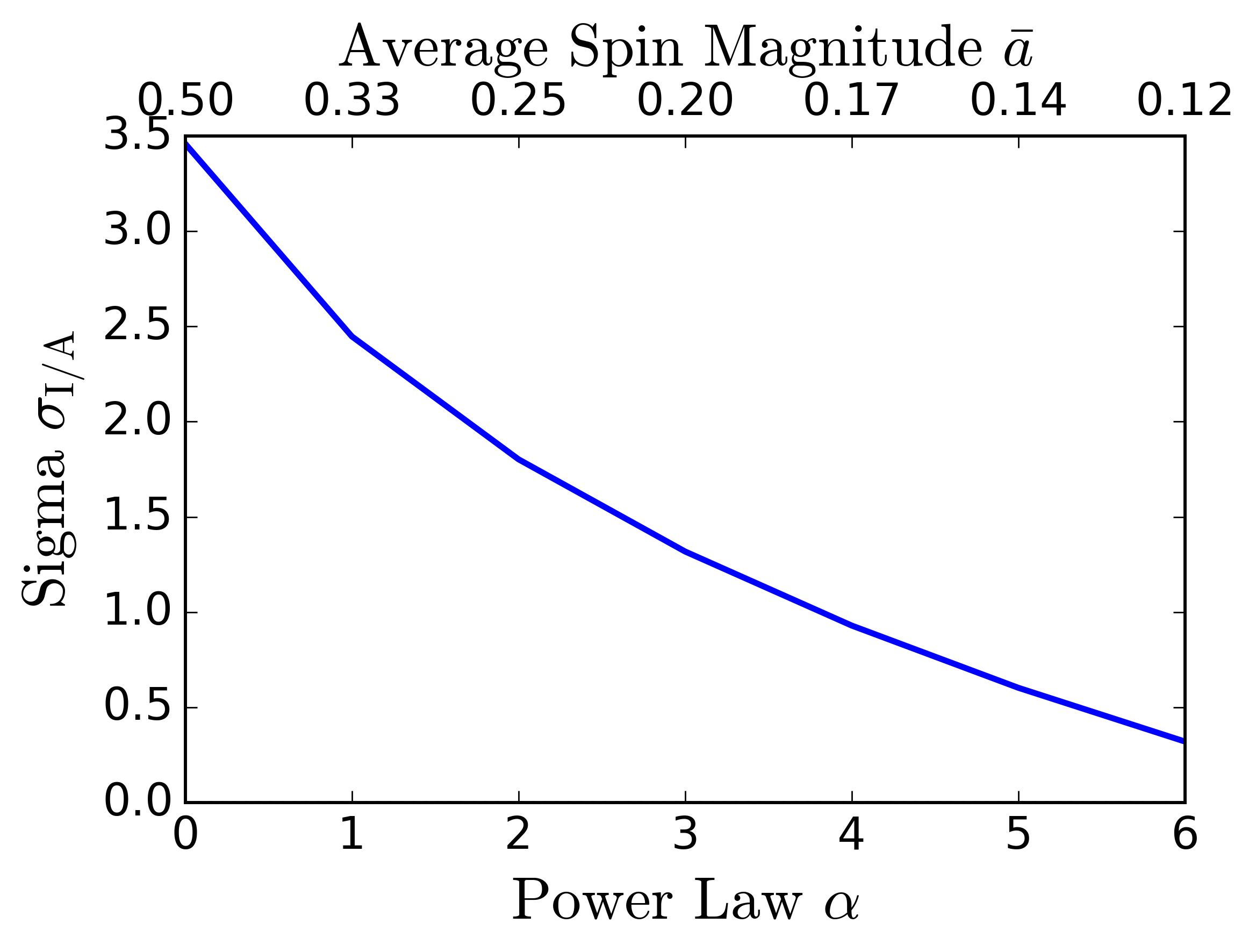}
\caption{\textbf{Effect of small spins on evidence ratio of isotropic against
  aligned models.} The blue line shows the evidence ratio
  (plotted as the equivalent sigma) between a model where all systems
  are from an isotropic distribution, versus one where all systems are
  aligned, as a function of the power law $\alpha$ corresponding
  to Equation~\ref{eq:lowspinalpha}.  The top axis shows the mean spin
  magnitude $\bar{a}$ which this $\alpha$ corresponds to. We see that
  for mean spin magnitudes $\lesssim 0.2$ we find no evidence for
  either distribution over the other.}
\label{fig:smallspinsalpha}
\end{figure}

In Extended Data Figure~\ref{fig:smallspinsalpha} we plot the evidence
ratio of isotropic to aligned distributions (plotted as the equivalent
sigma) with spin magnitudes given by this model with $\alpha$ in the
range $0$--$6$. The top axis shows the mean spin magnitude that value
of $\alpha$ corresponds to (e.g., for the ``flat'' distribution
$\alpha = 0$, the mean spin magnitude is 0.5). We see that if typical
BH spins are $\lesssim 0.2$ we have no evidence for one model over the
other.

\section{Mass Ratio}
\label{sec:mass-ratio}

Extended Data Figure \ref{fig:mass-ratio-sensitivity} shows the
distributions of $\chieff$ that would obtain with a mass ratio
$q = m_2/m_1 = 0.5$ compared to the distributions with $q = 1$ used
above.  The details of the distribution are sensitive to the mass
ratio, but in our analysis we are primarily sensitive to the changing
\emph{sign} of $\chieff$ under the isotropic models.  This latter
property is insensitive to mass ratio.  As an example, the distinction
between the three different spin amplitude distributions after ten
additional detections is quite weak compared to the aligned/isotropic
distinction in Figure \ref{fig:O2-predictions}.  The differences in
the $\chieff$ distribution between $q = 1$ and $q = 0.5$ are even
smaller than the differences between the different magnitude
distributions.

\begin{figure}
  \plotone{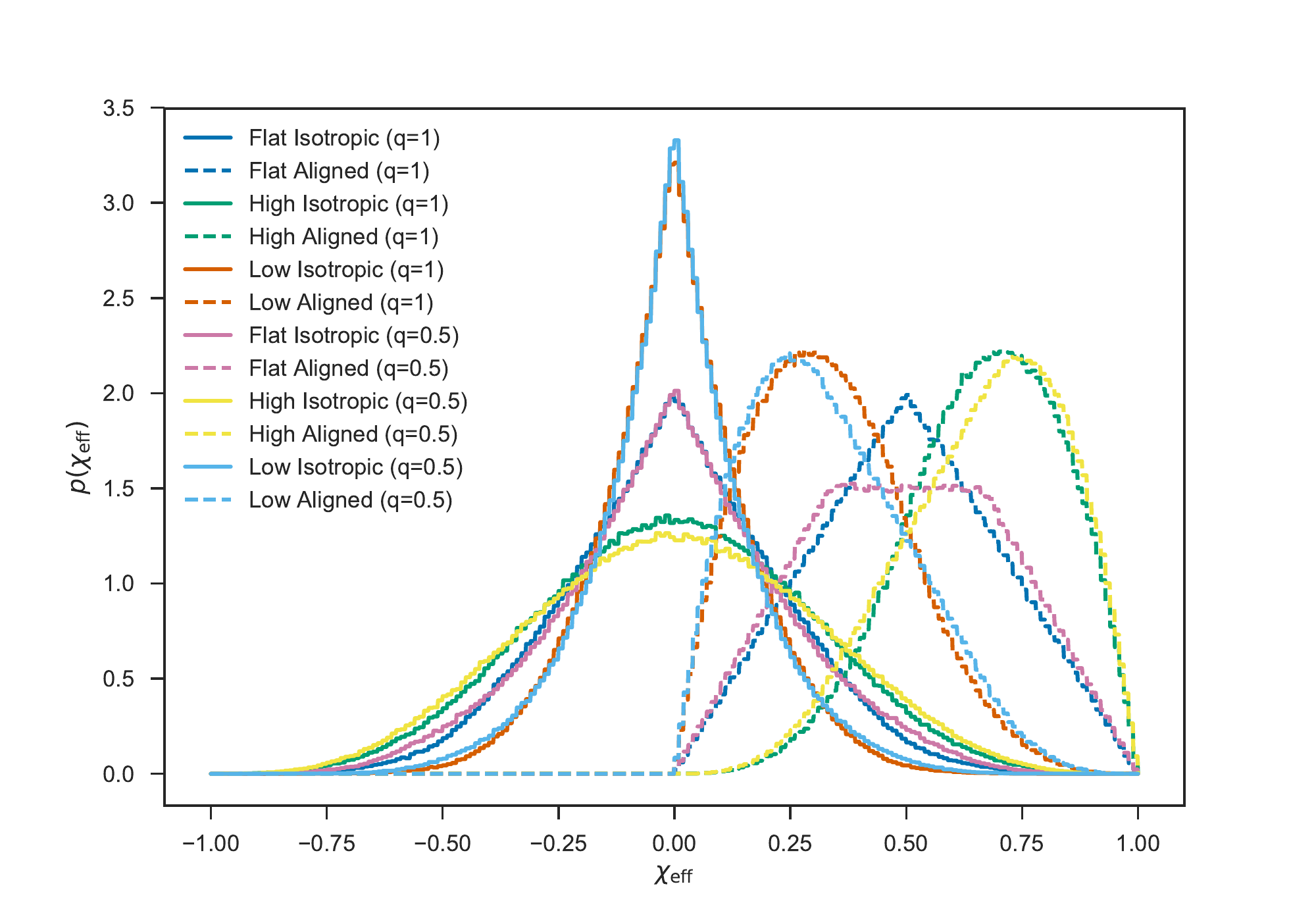}
  \caption{\textbf{Distributions of $\chieff$ assuming all merging
      black holes have equal masses ($q=1$) or a 2:1 mass ratio
      ($q = 0.5$).}  The details of the distribution are sensitive to
    the mass ratio, but in our analysis we are primarily sensitive to
    the changing \emph{sign} of $\chieff$ under the isotropic models.
    This latter property is unchanged under changing mass ratio.}
  \label{fig:mass-ratio-sensitivity}
\end{figure}

\section{Approximations in the Gravitational Waveform and Selection Effects}
While the Advanced LIGO searches use spin-aligned templates they are
efficient in detecting misaligned binary black hole
systems\cite{2016PhRvD..93l2003A}; we assume here that the $\chieff$
distribution of observed sources follows the true population.

The model waveforms used to infer the $\chieff$ of the three LIGO
events incorporate approximations to the true behaviour of the merging
systems that are expected to break down for sufficiently high
mis-aligned spins.  The effect of these approximations on inference on
the parameters describing GW150914 has been investigated in
detail\cite{2016arXiv161107531T}.  For this source, statistical
uncertainties dominate over any waveform systematics.  Detailed
comparisons with numerical relativity computations using no
approximations to the dynamics\cite{2016PhRvD..94f4035A} also suggest
that statistical uncertainties dominate the systematics for this
system.  Systematics may dominate for signals with this large SNR
($\simeq 23$) when the source is edge-on or has high
spins\cite{2016arXiv161107531T}.  The other two events discussed in
this paper are at much lower SNR, with correspondingly larger
statistical uncertainties, and are probably similarly oriented and
with similarly small spins, so we do not expect systematic
uncertainties to dominate.

We assume here that measurements made in the future are not dominated
by systematic errors, but this assumption would need to be revisited
for high-SNR, edge-on, or high-spin sources detected in the future.

\section{Precision of $\chieff$ measurements}
\label{sec:chi-eff-precision}
Throughout this work we have made the simplifying assumption that the
precision to which $\chieff$ can be constrained for individual binaries
is independent of the binary's properties.  In practice, our ability to
constrain $\chieff$ \emph{is} dependent on the system's properties, in
particular its true $\chieff$ and mass ratio, which we illustrate in
Extended Data Figure \ref{fig:chi-eff-constraints}.

For this figure a detected population\footnote{We qualify a system as
  ``detected'' if it produces a SNR above $8$ in the second-loudest
  detector to select only coincident events.} of 500 binaries was
selected from a population with component masses distributed uniformly
between $1$ and $30~\mathrm{M}_\odot$ with
$m_1 + m_2 < 30~\mathrm{M}_\odot$, locations distributed uniformly in
volume, and orientations distributed isotropically.  Data were
simulated for each binary, and posteriors were estimated using the
LIGO-Virgo parameter estimation library
\texttt{LALInference}\cite{2015PhRvD..91d2003V} using inspiral-only
waveform models (merger and ringdown effects can provide additional
information for some binaries, but we ignore those effects here).
$\chieff$ is better constrained for binaries with high
effective spins and high ($\sim$equal) mass ratios.

We do not expect these effects to qualitatively affect
out conclusions, though they could affect predictions for the total
number of detections necessary to constrain the population.  For
example, if the universe preferentially forms asymmetric binaries with
low mass ratios, individual $\chieff$ constraints will be systematically
worse, requiring more binaries to infer the properties
of the population.

\begin{figure}
  \plotone{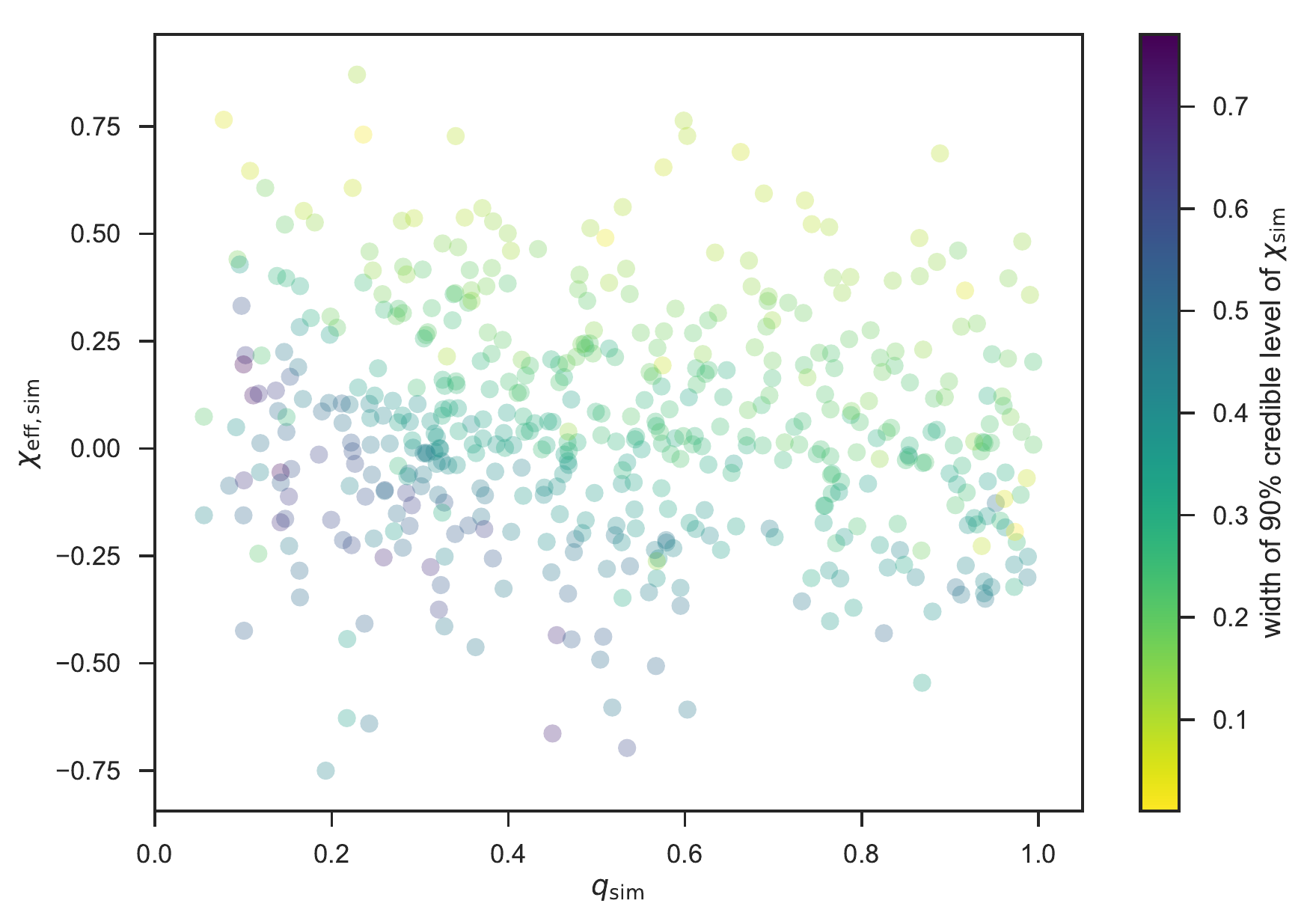}
  \caption{\textbf{Widths of the 90\% credible intervals for $\chieff$
      for 500 binaries in a simulated detected population.} $\chieff$
    is better constrained for systems with high $\chieff$ and high mass
    ratio.}
  \label{fig:chi-eff-constraints}
\end{figure}

\end{methods}

\bibliographystyle{naturemag-arxiv}
\bibliography{AlignedIsotropicGW}

\end{document}